\documentclass{iaus}

\usepackage{psfig}

\usepackage{graphics}

  \checkfont{eurm10}
  \iffontfound
    \IfFileExists{upmath.sty}
      {\typeout{^^JFound AMS Euler Roman fonts on the system,
                   using the 'upmath' package.^^J}%
       \usepackage{upmath}}
      {\typeout{^^JFound AMS Euler Roman fonts on the system, but you
                   dont seem to have the}%
       \typeout{'upmath' package installed. iaus.cls can take advantage
                 of these fonts,^^Jif you use 'upmath' package.^^J}%
      }
  \else
  \fi

  \checkfont{msam10}
  \iffontfound
    \IfFileExists{amssymb.sty}
      {\typeout{^^JFound AMS Symbol fonts on the system, using the
                'amssymb' package.^^J}%
       \usepackage{amssymb}%

      }{}
  \fi

  \IfFileExists{amsbsy.sty}
    {\typeout{^^JFound the 'amsbsy' package on the system, using it.^^J}%
     \usepackage{amsbsy}}
    {}

\newsavebox{\astrutbox}
\sbox{\astrutbox}{\rule[-5pt]{0pt}{20pt}}

\newcommand\etal{\mbox{\textit{et al.}}}

\def\halfls{\vskip 6pt}                                
\def\ni{\noindent}                                            
\def\etal{{\it et\thinspace al.}\ }                              
\def\lapprox{$_<\atop{^\sim}$}            

\title[The Interplay among Black Holes, Stars and ISM in Galactic 
       Nuclei]{Stellar Populations in Active Galaxies}

\author[R. Cid Fernandes]%
{R. Cid Fernandes$^1$}

\affiliation{$^1$Universidade Federal de Santa Catarina,
Florian\'opolis, SC, Brazil email: cid@astro.ufsc.br}

\pubyear{2004}
\volume{222}
\pagerange{1--8}
\date{?? and in revised form ??}
\jname{The Interplay among Black Holes, Stars and ISM \\in Galactic Nuclei}
\editors{Th. Storchi Bergmann, L.C. Ho \& H.R. Schmitt, eds.}
\begin{document}

\maketitle

\begin{abstract}
The role of stars and starbursts in AGN has been a recurring issue for
nearly as long as AGN have been recognized as hosts of interesting
phenomena.  The heated ``starburst {\it versus} monster'' controversy
of the 80's and 90's was gradually replaced by ``starburst {\it plus}
monster'' studies, as observational work in the past decade has firmly
established that accretion onto a super-massive black-hole and
star-formation coexist in many galactic nuclei.  Whereas the physical
link between starbursts and AGN remains unclear, there remains no
doubt that starbursts affect a number of properties traditionally
associated to the AGN alone, such as the so called ``featureless
continuum'', emission line ratios and luminosities.  This contribution
glosses over some of the techniques used to diagnose stellar
populations in AGN, focusing on recent results and how this type of
work can lead us well beyond what became known as the starburst-AGN
connection.
\end{abstract}

\firstsection 
\section{Introduction}

The subject of stellar populations in AGN has gone through a history
of love and hate over the past 3 decades. Once upon the time, stars
were essentially seen as that unavoidable junk which pollutes the
optical spectra of AGN, particularly type 2s (Seyfert 2s, LINERs and
their relatives).  Accordingly, the methodology to deal with starlight
in those days was philosophically the same used to deal with sky
features or cosmic rays: {\it Get rid of it!} This was achieved by
modeling the spectrum as a combination of an elliptical galaxy
template plus a non-stellar featureless continuum (FC), yielding the
``pure-AGN'' spectrum as the residual (Koski 1978). Since this
approach postulates that stars in AGN are all old and boring, it is
not surprising that few AGNauts cared about stellar populations at
all, so meetings like these would not have been possible at the time.
This ``Get Rid of It'' era lasted up to the mid-80's.

In the late 80's and 90's things changed radically from a
``stars-have-nothing-to-do-with-AGN'' to the ``stars = AGN'' idea put
forward by R. Terlevich and collaborators, whose starburst model for
AGN replaced super-massive black-holes (SMBHs) by young stars and
their remnants as the power-engine of active galaxies.  Those of use
which were around at the time remember (some with nostalgia) the fierce
debates in meetings across the globe (eg, Taipei 1992, Puebla
1996). The conflicting observational evidence back then fueled the
controversy. For instance, while the discovery of Seyfert 1-like SNe
supported the model, rapid X-ray variability and relativistic Fe
K$\alpha$ line-profiles were better understood in the framework of the
black-hole paradigm.

The starburst $\times$ monster battle gradually disappeared from the
headlines as the existence of SMBHs became conclusively established in
the past decade. The interest in stellar populations in AGN, however,
survives to the present day, and for a very good reason: Evidence of
the presence of young and intermediate age population around AGN is
now as solid as that for the existence of SMBHs! Young ($t < 10^7$ yr)
and ``mature'' ($10^{8 {\rm -} 9}$ yr) starbursts have been found all
across the AGN family: quasars (Brotherton \etal 1999), radio galaxies
(Wills \etal 2002), Seyfert 2s (Heckman \etal 1997) and LLAGN (Cid
Fernandes \etal 2004; Gonz\'alez-Delgado \etal 2004). This very volume
reports new developments on this so called ``starburst-AGN
connection'', as in the contributions by Rafaela Morganti, Wil van
Breugel, Rosa Gonz\'alez Delgado and others.  Indeed, the mere fact
that the word ``stars'' figures alongside ``black-holes'' in the title
of an IAU symposium is a testimony of this new reality.

Having said that, I must point out that we are still a long way away
from understanding the physical link between star-formation and SMBHs
in AGN. In fact, except for a few cases (see van Breugel's
contribution) we are not even sure such a causal link exists at all,
as the coexistence of these two phenomena could simply reflect the
fact that both live on the same gas-based diet, a trivial possibility
one must always bear in mind.

Notwithstanding this obligatory warning, we have learned a great deal
from such studies.  Instead of attempting a thorough and fair review
of the progress in the field, which would be impossible due to lack of
space and mainly talent, these few pages highlight some recent results
which illustrate how the careful modeling of stellar populations in
AGN provides valuable tools both for researchers interested in
studying starburst-AGN connections and those more interested in
getting rid of the starlight to inspect the AGN itself. The work cited
below was carried out with several friends, including R. Gonz\'alez
Delgado, H. Schmitt, T. Heckman, L. Martins, Q. Gu, J. Melnick, the
Terlevichs, D. Kunth, my students and, last but not least, our
chair-woman T. Storchi-Bergmann, who taught me a lot and worked very
hard to organize this great meeting.

\section{The tools of the trade: Diagnosing stellar population in AGN}

\subsection{Spectral features}

\label{sec:SpectralFeatures}

Stars leave numerous imprints in an AGN spectrum. These are more
easily detected when the AGN is weak or hidden, which explains why
most of the progress in this field has been achieved studying type 2
nuclei. An incomplete inventory of stellar features detected in AGN
include the PCygni lines of CIV and SiIV in the UV (Gonz\'alez Delgado
\etal 1998), the WR bump at $\sim 4650$ \AA, high order Balmer
absorption lines (Storchi-Bergmann \etal 2000) , the CaII K, G-band
and MgI absorptions in the optical, the CaII triplet (Terlevich,
D\'{\i}az \& Terlevich 1990) and Si and CO bands in the near-IR (Oliva
\etal 1999). Star-related features are also found, such as a soft
component in X-rays due to a starburst driven wind (Levenson, Weaver
\& Heckman 2001). Another example is the 3.3 $\mu$m PAH feature
associated with starburst activity (Rodr\'{\i}guez-Ardila \& Viegas
2003), which, unlike most other tracers, can be applied to type 1 AGN
and thus help solving the old conundrum that starbursts seem to be
less frequent in Seyfert 1s than in Seyfert 2s, in contradiction with
the unification paradigm.

Some of these features are easily transformed into stellar population
diagnostics, while others require more work. For instance, the mere
detection of PCygni lines or the WR bump implies the presence of
starbursts a few Myr old, and Balmer lines signal the presence of
0.1--1 Gyr populations. The CaII K line and the 4000 \AA\ break, on
the other hand, can be diluted either by young stars or by AGN-light
(the ``starburst-FC degeneracy''), so their interpretation is
trickier.  The Ca triplet and CO lines in the near-IR by themselves do
not offer great diagnostic power, but can be used to infer the stellar
kinematics and the $M/L$ ratio (Oliva \etal 1999), a tool which is yet
to be fully explored in AGN.

\subsection{The new generation of stellar population tools}

\begin{figure}
\psfig{file=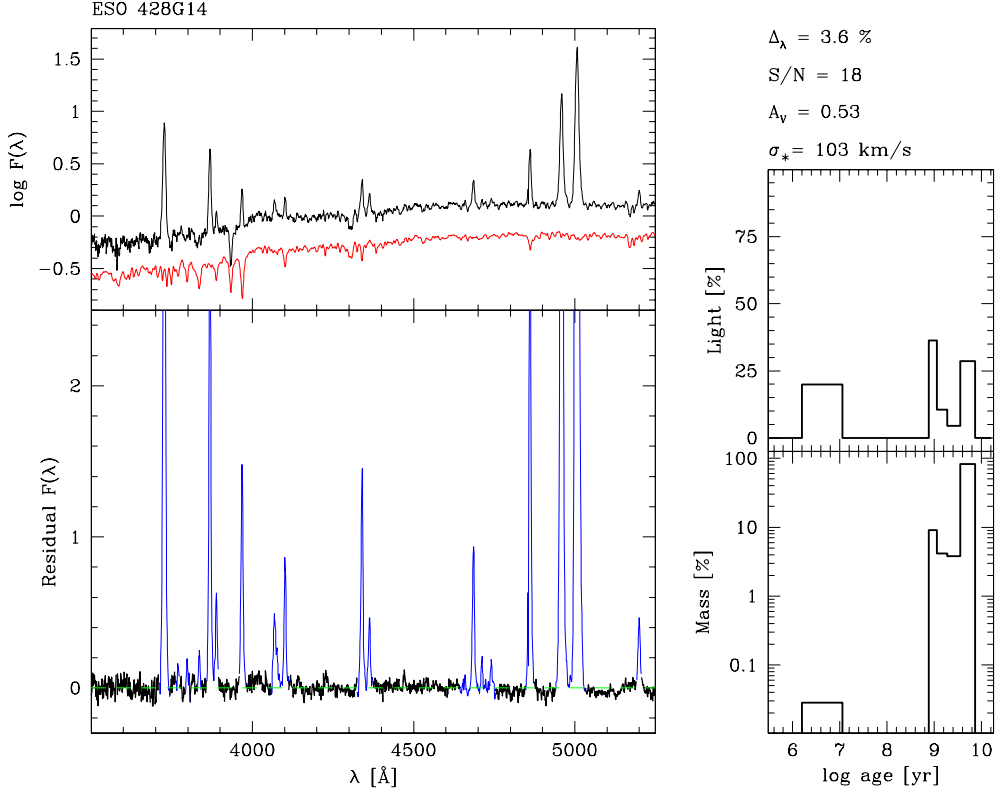,width=14.3cm,rheight=11cm}
\caption{Example of the accurate spectral fits obtained with the new
generation of stellar population synthesis models. {\it Left:}
Observed, model and residual spectra. {\it Right:} Star-formation
history, expressed as light and mass fractions versus age.}
\label{fig:specfit}
\end{figure}

The release of high spectral resolution evolutionary synthesis models
by Bruzual \& Charlot last year provided a much awaited tool to study
stellar populations in galaxies. As other groups, we have implemented
this new library in a code which fits a galaxy spectrum decomposing it
into $N_\star$ simple stellar populations of different ages and
metallicities. It so happened that at the time we were starting to
analyze the stellar populations of a large and homogeneous sample of
$\sim 70$ Seyfert 2s, which sounded like a good application for our
code. The resulting spectral fits were incredibly good, as illustrated
in Fig~\ref{fig:specfit}.

\begin{figure}
\resizebox{7.2cm}{!}{\includegraphics{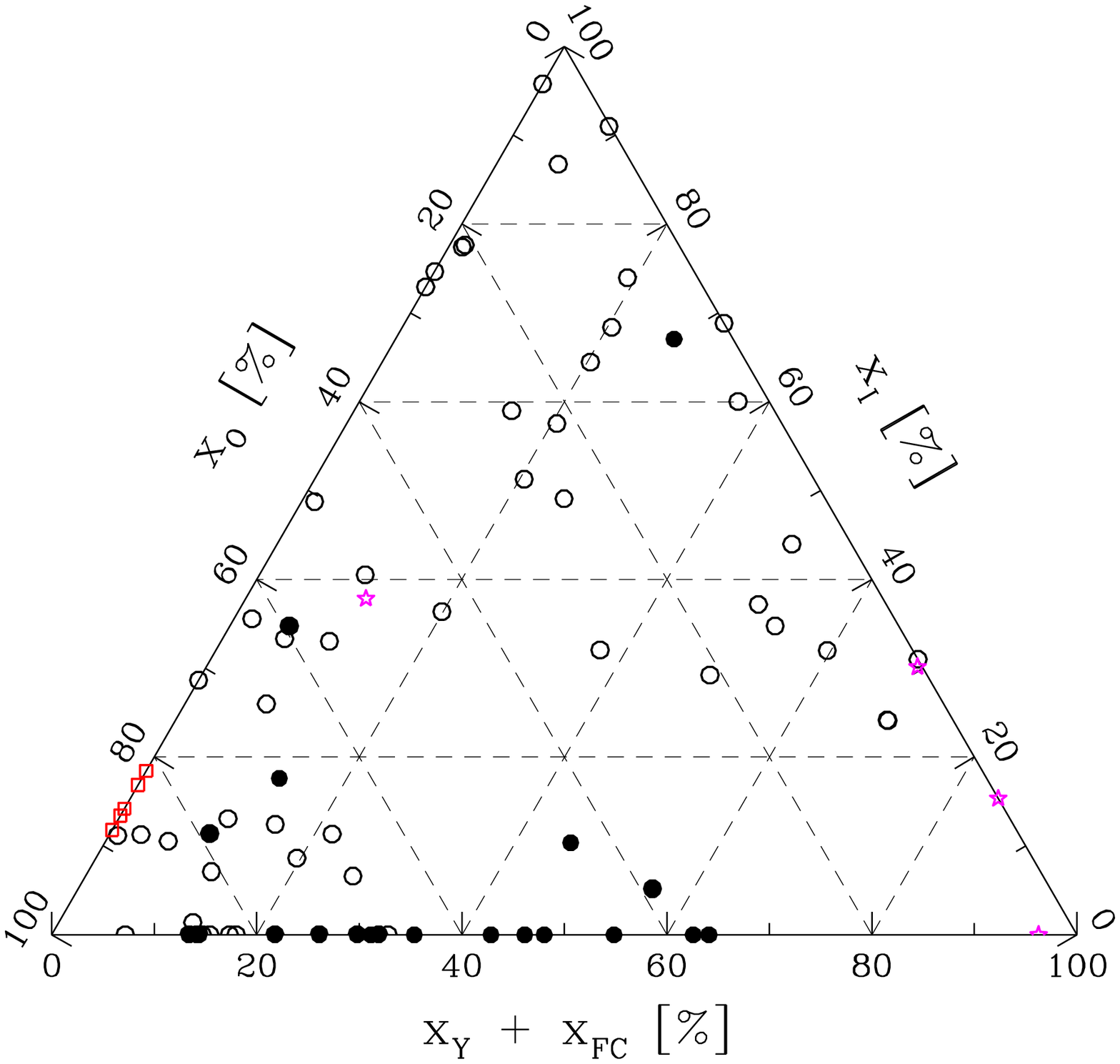}}
\resizebox{6.4cm}{!}{\includegraphics{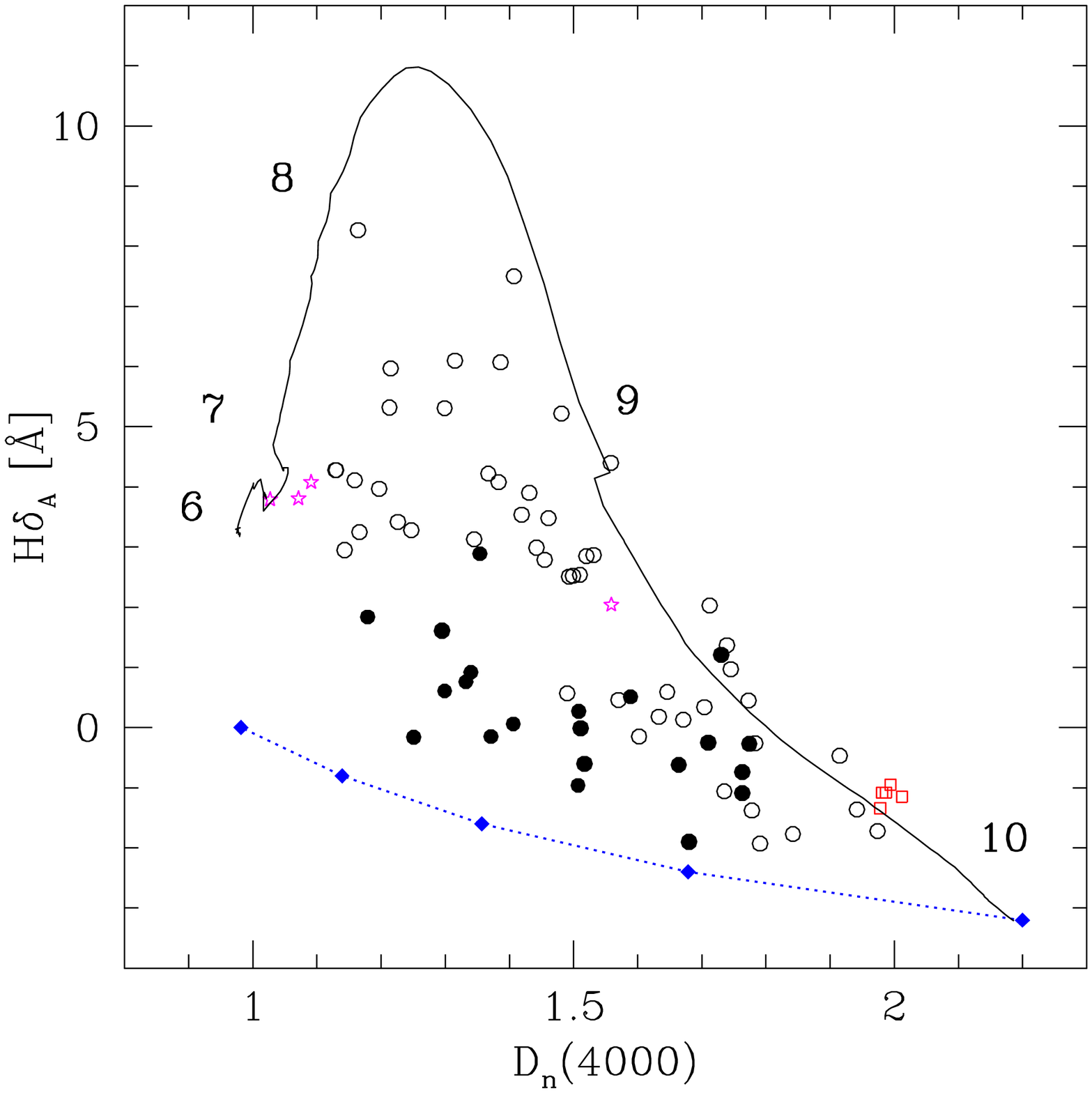}}
\caption{{\it Left:} Light fraction (at $\lambda = 4020$ \AA)
associated to Young stars plus an FC ($x_Y + x_{FC}$), Intermediate
age ($x_I$) and Old populations ($x_O$). {\it Right:} $D_n(4000)$
versus H$\delta_A$ diagram. The solid line shows the evolution of a
BC03, $Z_\odot$ instantaneous burst, with numbers indicating ages of
$10^{6 \ldots 10}$ yr. The dashed line represents a mixture of a
$F_\nu \propto \nu^{-1.5}$ FC with a $10^{10}$ yr
elliptical-galaxy-like stellar population.  Circles represent the 65
Seyfert 2s in this sample; filled circles mark ``Broad Line Seyfert
2s'', in which scattered light is detected.}
\label{fig:HdD4000}
\end{figure}

Fig \ref{fig:HdD4000} summarizes the properties of the central stellar
populations of nearby Seyfert 2s. The left panel shows the population
vector divided into ``Young'' (+ FC), ``Intermediate age'' and ``Old''
age-bins. The right plot expresses essentially the same information
but in the more empirical $H_\delta \times D_n(4000)$ diagram. Seyfert 2s
are scattered all over these diagrams, implying a wide variety of
nuclear stellar population properties. At first sight this diversity
seems to suggest no connection between their AGN and the stars around
it. However, the $\sim 40\%$ frequency of \lapprox 1 Gyr populations
in Seyfert 2s is high compared to galaxies of the same (early) Hubble
type, and there are important connections between such starbursts and
the AGN luminosity (Kauffman \etal 2003).

This new generation of stellar populations synthesis superseeds
diagnostics based on individual spectral indices
(\S\ref{sec:SpectralFeatures}), as modeling the full spectrum yields
much better constrained fits and produces far more informative output
information. Stellar masses, velocity dispersions (which gained added
value in light of the $M_{\rm BH}$-$\sigma_\star$ relation which
occupies several pages of this volume), extinction and the whole
star-formation history of the system can be derived!  As if this
wasn't enough, to our surprise we are finding that even {\it stellar
metallicities} can also be recovered within reasonable uncertainties.
Applications of such techniques to large AGN samples are starting to
appear. Two important results of this kind of study are that AGN live
almost exclusively within massive galaxies, and powerful AGN tend to
have younger stellar populations (Kauffman \etal 2003). We have just
started to play with SDSS galaxies, and, besides confirming Kauffman's
results, we also find that AGN have metal rich stellar populations.

\section{So What?}

Given the lingering doubts about whether there is an actual physical
connection between starbursts and AGN, I find it relevant to point out
that modeling stellar populations in AGN has a lot to offer to those
which don't care at all about stars, but, in the spirit of the ``Get
Rid of It'' era, must deal with them. Here are a few why's.

\subsection{Stellar populations as clocks to measure AGN evolution}

A central goal of stellar population synthesis is to produce {\it age}
estimates ($t_\star$). Knowing the ages of stars around AGN can give
us major clues as to the evolution of AGN themselves! The idea is to
age-date as many and as diverse AGN as we can and check whether their
observed (luminosities, line-ratios, etc) and physical properties
($M_{BH}$, $\dot{M}_{BH}$) evolve, ie, if they correlate with
$t_\star$. We are still in the first stages of this process, but one
result is already clear from our studies of Starburst galaxies,
Seyfert 2s, Transition Objects (TOs) and LINERs: On average,

\halfls \centerline{$\overline{t_\star}$(Starbursts) $<
\overline{t_\star}$(Seyfert 2s) $< \overline{t_\star}$(Young TOs) 
 $< \overline{t_\star}$(Old TOs) $=
\overline{t_\star}$(LINERs)} \halfls

\ni where Young/Old refers to the presence of detectable of \lapprox 1
Gyr-old populations---as discussed by Gonz\'alez Delgado elsewhere in
this volume, Young LINERs don't seem to exist! Now, before jumping to
interpretations recall that we are comparing apples and pineapples
insofar as Hubble types are concerned. Seyferts and LINERs live in
early types galaxies, Starbursts live in late types, with TOs split in
between (Ho, Filippenko \& Sargent 2003). The two plausible
evolutionary sequences are thus: Seyfert 2 $\rightarrow$ Young TO
$\rightarrow$ Old TO/LINER, and Starburst $\rightarrow$ Young TO
$\rightarrow$ Old TO. While appealing, this is not really the whole
story, since at least 30\% of Seyfert 2s look just like Old TOs and
LINERs in their stellar populations. Though much remains to be done,
this seems a promising vein to explore, particularly in these days of
mega-surveys.

\subsection{Detection of scattered light in Seyfert 2s}

\begin{figure}
\psfig{file=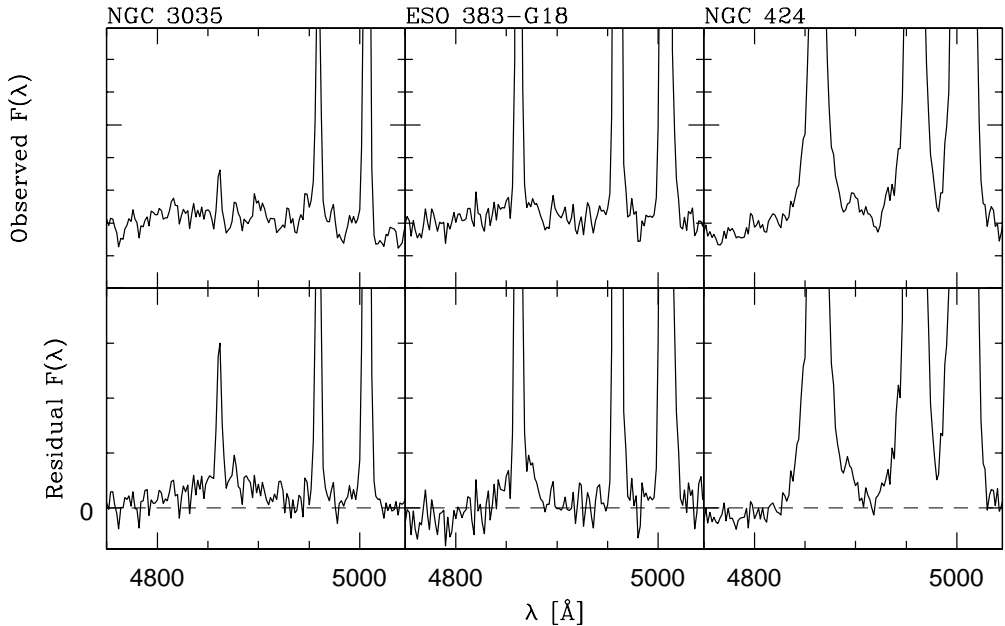,width=14.2cm,rheight=8.8cm}
\caption{Top: Observed H$\beta$ profiles in three Seyfert 2s. Bottom:
H$\beta$ profiles after subtracting the model spectrum, revealing a
BLR-like component which we interpret as scattered light.}
\label{fig:BLRSey2s}
\end{figure}

Even the most starburst-AGN-connection-skeptic would agree that the
elliptical galaxy $+$ FC equation of the 70's and 80's is horribly
wrong for many Seyfert 2s. One consequence of this template mismatch
is that it often produced too strong an FC to be compatible with the
unified model given absence of broad lines in Seyfert 2s and their low
polarizations (Cid Fernandes \& Terlevich 1995). This gave rise to the
so called ``FC2-problem'' (Tran 1995). It is now clear that these 2nd
(and dominant) FC component is originated in starbursts.  Up to very
recently the scattered Seyfert 1 component (called ``FC1'' by the
pundits) seen via spectropolarimetry remained too weak to be detected
in direct light.

This is no longer true.  The new synthesis models do such a good job
that we are now capable of spotting weak broad lines in the observed
$-$ model residual spectra, as shown in Fig \ref{fig:BLRSey2s}. Our
stars plus power-lawish FC spectral fits find significant FC strengths
for most of these ``Broad Line Seyfert 2s'', as well as for objects
where a hidden Seyfert 1 is known to exist from spectropolarimetry
work (filled circles in Fig \ref{fig:HdD4000}). The conclusion is
obvious: We are detecting scattered light without a polarimeter! This
is just an illustration of what population synthesis can do for you:
The superb fits provided by these tools produce a clean ``pure-AGN''
spectrum, revealing weak features which would otherwise be hard to
detect. This applies both to continuum and line emission.

\subsection{Dust and intermediate age stars in Low Luminosity AGN}

The third word in our meeting's title is ISM, of which dust is a
part. Here too stellar population synthesis has a saying. For
instance, in a spatially resolved analysis of LLAGN spectra, we have
just found that Young TOs are much more reddened than either Old TOs
or LINERs, and that this dust is concentrated in the nuclear
regions. This dust could be due to star-forming regions, in analogy
with what is found in starburst + Seyfert 2 composites, which tend to
be dustier than Seyfert 2s whose inner stellar populations are
predominantly old. However, no obvious signs of young ($< 10^7$ yr)
starbursts are evident in most TOs, contrary to most expectations (eg,
Shield's and Sarzi's talks).  Alternatively, this dust could be
associated to post-AGB stars, notorious dust-producers which have the
right Gyr-scale age-range to match the observed spectra and whose
ionizing radiation is also capable of producing TO-like emission lines
(Binette \etal 1994).

\section{Final remarks}

I hope these pages have conveyed the idea that we have learned (the
hard way, as usual) that the issue is not stars {\it or} monsters, but
stars {\it and} monsters, and specially that the stellar content
within the vicinity of the monster contains useful information for AGN
studies.  Historical difficulties in retrieving this information from
data are quickly being overcome by a new generation of powerful
stellar population synthesis tools, whose application to data of ever
increasing quantity and quality is doomed to shed new light into our
understanding of these fascinating objects.

\end{document}